\documentstyle[preprint,aps,epsf]{revtex}
\begin{document}
\draft
\preprint{}
\title{Structure of the Vacuum in Nuclear Matter --- A Nonperturbative Approach}
\author{A. Mishra$^{*\dagger}$, P.K. Panda\footnote[1]{Alexander von Humboldt
Fellows (email: mishra@th.physik.uni-frankfurt.de)}$^\dagger$, 
S. Schramm$^\ddagger$, J. Reinhardt$^\dagger$, 
W. Greiner$^\dagger$}
\address{$^\dagger$Institut f\"ur Theoretische Physik, 
J.W. Goethe Universit\"at, Robert Mayer-Stra{\ss}e 10,\\
Postfach 11 19 32, D-60054 Frankfurt/Main, Germany}
\address{$^\ddagger$ Gesellschaft f\"ur Schwerionenforschung (GSI),
Planckstra{\ss}e 1,\\ Postfach 110 552, D-64220 Darmstadt, Germany}
\maketitle
\begin{abstract}
We compute the vacuum polarisation correction to the binding energy
of nuclear matter in the Walecka model using a nonperturbative approach.
We first study such a contribution as arising from a ground state
structure with baryon-antibaryon condensates. This yields the same
results as obtained through the relativistic Hartree approximation
of summing tadpole diagrams for the baryon propagator. Such a vacuum 
is then generalized to include quantum effects from meson fields
through scalar-meson condensates. The method is applied to study
properties of nuclear matter and leads to a softer equation of
state giving a lower value of the incompressibility than would be
reached without quantum effects. The density dependent effective 
sigma mass is also calculated including such vacuum polarisation effects.   
\end{abstract}

\bigskip
\narrowtext

\section{Introduction}

Quantum Hadrodynamics (QHD) is a general framework for the nuclear
many-body problem \cite{qhd,reinhard,serot2}. It is a renormalisable
relativistic quantum field theory using hadronic degrees of freedom
and has quite successfully described the properties of nuclear matter and 
finite nuclei. In the Walecka model (QHD-I) with nucleons interacting with
scalar ($\sigma$) and vector ($\omega$) mesons, it has been shown
in the mean-field approximation that the saturation density and 
binding energy of nuclear matter may be fitted by adjusting the scalar
and vector couplings \cite{walecka}.
This was first done by neglecting the Dirac sea and is  called
the no-sea approximation.  In this approximation,
several groups have investigated
the effects of scalar self-interactions in nuclear matter \cite{nm}
and finite nuclei \cite{finnl} using a  mean-field approach. 

To include the sea effects, one does
a self-consistent sum of tadpole diagrams for the baryon
propagator \cite{chin}. This defines the relativistic Hartree
approximation. There have also been calculations including
corrections to the binding energy up to two-loops \cite{twoloop},
which are seen to be rather large as compared to the one-loop
results. However, it is seen that using phenomenological monopole form
factors to account for the composite nature of the nucleons,
such contribution is reduced substantially \cite{prakash}
so that it is smaller than the one-loop result.
Recently, form factors have been introduced as a cure to the
unphysical modes, the so called Landau poles \cite{horowitz},
which one encounters while calculating the meson propagator
as modified by the interacting baryon propagator of relativistic
Hartree approximation. There  have been also attempts to calculate
the form factors by vertex corrections \cite{allen}.
However, without inclusion of such form factors
the mean-field theory is not stable 
against a perturbative loop expansion. This might be because
the couplings involved here are too large (of order of 10)
and the theory is not asymptotically free. Hence 
nonperturbative techniques need to be developed to consider
nuclear many-body problems. The present work is a step in 
that direction including vacuum polarisation effects.

The approximation scheme here uses a squeezed coherent type of 
construction for the ground state \cite{cond,ph4} which amounts 
to an explicit vacuum realignment. The input here is
equal-time quantum algebra for the field operators
with a variational ansatz for the 
vacuum structure and does not use any perturbative expansion
or Feynman diagrams. We have earlier seen that this correctly
yields the results of the Gross-Neveu model \cite{gn} as obtained 
by summing an infinite series of one-loop diagrams. We have also
seen that it reproduces the gap equation in an effective QCD
Hamiltonian \cite{chirl} as obtained through the solution of the
Schwinger-Dyson equations for the effective quark propagator.
We here apply such a nonperturbative 
method to study the quantum vacuum in nuclear matter.

We organise the paper as follows. In section 2, 
we study the vacuum polarisation effects in nuclear matter
as simulated through a vacuum realignment with baryon-antibaryon condensates.
The condensate function is determined through a minimisation
of the thermodynamic potential. 
The properties of nuclear matter as arising from such a vacuum
are then studied and are seen to become
identical to those obtained through the relativistic Hartree approximation.
In section 3, we generalise such a vacuum state to include sigma 
condensates also, which are favoured with a quartic term 
in the sigma field in the Lagrangian. The quartic coupling 
is chosen to be positive which is necessary 
to consider vacuum polarisation effects from the sigma field.
We also calculate the effective sigma mass arising through such quantum
corrections as a function of density. The coupling here  
is chosen to give the value for the incompressibilty of
nuclear matter in the correct range. Finally, in section 4,
we summarise the results obtained through our nonperturbative approach
and present an outlook.

\section {Vacuum with baryon and antibaryon condensates}
We start with the Lagrangian density for the linear Walecka model given as
\begin{eqnarray}
{\cal L}=\bar \psi (i\gamma^\mu \partial_\mu
-M-g_\sigma \sigma-g_\omega\gamma^\mu \omega_\mu)\psi\nonumber\\
&+&\frac{1}{2}\partial^\mu\sigma
\partial_\mu\sigma-\frac{1}{2} m_\sigma ^2 \sigma^2
+\frac{1}{2} m_\omega^2 \omega^\mu \omega_\mu
-\frac{1}{4}\omega^{\mu \nu}\omega_{\mu \nu},
\end{eqnarray}
with
\begin{equation}
\omega_{\mu \nu}=\partial_\mu \omega_\nu -\partial_\nu \omega_\mu.
\end{equation}
In the above, $\psi$, $\sigma$, and $\omega_\mu$ are the fields
for the nucleon, $\sigma$-, and $\omega$-mesons
with masses $M$, $m_\sigma$, and $m_\omega$, respectively. 

We use the mean-field approximation for the meson  
fields and retain the quantum nature of the fermion fields \cite{gn}.
This amounts to taking meson fields as constant classical
fields with translational invariance for nuclear matter. Thus we shall replace
\begin{mathletters}
\begin{equation}
g_\sigma \sigma\rightarrow\langle g_\sigma \sigma\rangle
\equiv g_\sigma \sigma_0
\end{equation}
\begin{equation}
g_\omega \omega_\mu\rightarrow \langle g_\omega \omega_\mu\rangle
\equiv g_\omega \omega_\mu\delta^{\mu 0}=g_\omega \omega_0 
\end{equation}
\end{mathletters}
where $\langle \cdots\rangle$ denotes the expectation value in 
nuclear matter and we have retained the zeroth component for 
the vector field to have nonzero expectation value.

The Hamiltonian density can then be written as
\begin{equation}
{\cal H}={\cal H}_N+{\cal H}_\sigma+{\cal H}_\omega
\end{equation}
with
\begin{mathletters}
\begin{equation}
{\cal H}_N=\psi ^\dagger(-i \vec \alpha \cdot \vec \bigtriangledown
+\beta M)\psi + g_\sigma\sigma \bar \psi\psi
\end{equation}
\begin{equation}
{\cal H}_\sigma= \frac{1}{2} m_\sigma ^2 \sigma^2
\label{lwsg}
\end{equation}
\begin{equation}
{\cal H}_\omega= g_\omega \omega_0 \psi
^\dagger \psi
 -\frac{1}{2} m_\omega ^2 \omega_0^2
\end{equation}
\end{mathletters}
The equal-time quantization condition for the nucleons is given as
\begin{equation}
[\psi _{\alpha}(\vec x,t),\psi^\dagger _\beta (\vec y,t)]_{+}
=\delta _{\alpha \beta}\delta(\vec x -\vec y),
\end{equation}
where $\alpha$ and $\beta$ refer to the spin indices.
We may now write down the field expansion for the nucleon field
$\psi$ at time $t=0$ as given by \cite{prd}
\begin{equation}
\psi(\vec x)=\frac {1}{(2\pi)^{3/2}}\int \left[U_r(\vec k)c_{Ir}(\vec k)
+V_s(-\vec k)\tilde c_{Is}(-\vec k)\right] e^{i\vec k\cdot \vec x} d\vec k,
\end{equation}
with $c_{Ir}$ and $\tilde c_{Is}$ as the baryon  annihilation 
and antibaryon creation operators with spins $r$ and $s$, respectively.
In the above, $U_r$ and $V_s$ are given by

\begin{equation}
U_r(\vec k)=\left( \begin{array}{c}\cos\frac{\chi(\vec k)}{2}
\\ \vec \sigma \cdot\hat k\sin\frac{\chi(\vec k)}{2}
\end{array}\right)u_{Ir} ;\quad V_s(-\vec k)=\left(
\begin{array}{c}-\vec \sigma \cdot\hat k\sin\frac{\chi(\vec k)}{2}
\\ \cos\frac{\chi(\vec k)}{2}\end{array}\right)v_{Is},
\end{equation}
For free massive fields $\cos\chi(\vec k)=M/\epsilon(\vec k)$ and
$\sin\chi(\vec k)=|\vec k|/\epsilon(\vec k),$ with 
$\epsilon(\vec k)=\sqrt{\vec k^2 + M^2}$.

The above are consistent with the equal-time anticommutator
algebra for the operators $c$ and $\tilde c$ as given by
\begin{equation}
[c_{Ir}(\vec k),c_{Is}^\dagger(\vec k')]_{+}=
\delta _{rs}\delta(\vec k-\vec k')=
[\tilde c_{Ir}(\vec k),\tilde c_{Is}^\dagger(\vec k')]_{+}.
\end{equation}
The perturbative vacuum, say $\mid vac\rangle$, is defined through
$c_{Ir}(\vec k)\mid vac\rangle=0$ and $\tilde c_{Ir}^\dagger(\vec k)\mid 
vac \rangle=0$.

To include the vacuum-polarisation effects, we shall now consider a 
trial state with baryon-antibaryon condensates. We thus explicitly
take the ansatz for the above state as 
\begin{eqnarray}
|vac'\rangle &=& \exp \Big[
\int d\vec k ~f(\vec k)~{c_{Ir}^\dagger (\vec k)}~a_{rs}
\tilde c _{Is} (-\vec k)-h.c. \Big]|vac\rangle \nonumber\\
& \equiv & U_F|vac\rangle,
\label{barcon}
\end{eqnarray}
Here $a_{rs}=u_{Ir}^\dagger(\vec \sigma \cdot \hat k)v_{Is}$ 
and $f(\vec k)$ is a trial function associated with baryon-antibaryon 
condensates. We note that with the above transformation
the operators corresponding to $\mid vac'\rangle$ are related to the
operators corresponding to $|vac\rangle$ through the Bogoliubov transformation
\begin{equation}
\left(
\begin{array}{c} d_{I}(\vec k)\\{\tilde d}_{I}(-\vec k)
\end{array}
\right)
=\left(
\begin{array}{cc}
\cos f(\vec k) & -\vec\sigma \cdot \hat k\sin f(\vec k)
\\ \vec\sigma \cdot \hat k \sin f(\vec k) & \cos f(\vec k)
\end{array}
\right)
\left(
\begin{array}{c}
c_I(\vec k)\\ {\tilde c}_I(-\vec k)
\end{array}
\right ),
\label{chi8}
\end{equation}
for the nucleon.

We then use the method of thermofield dynamics \cite{tfd}
developed by Umezawa to construct the ground state for nuclear matter. 
We generalise the state with baryon-antibaryon condensates
as given by (\ref{barcon}) to finite temperature and density as
\cite{ph4}
\begin{equation}
|F(\beta)\rangle=U(\beta)|vac'\rangle\equiv U(\beta)U_F|vac\rangle.
\end{equation}
The temperature-dependent unitary operator $U(\beta)$ is given as
\cite{tfd}
\begin{equation}
U(\beta) =\exp{(B^\dagger (\beta)-B(\beta))},
\end{equation}
with 
\begin{equation}
B^\dagger(\beta)= \int d\vec k ~\bigg[\theta_-(\vec k,\beta)~
d_{Ir}^\dagger (\vec k)~{\underline d}_{Ir}^\dagger(-\vec k)
+\theta_+(\vec k,\beta)~
\tilde d_{Ir}(\vec k)~\tilde {\underline d}_{Ir}(-\vec k)\bigg].
\end{equation}
The underlined operators are the operators corresponding to the
doubling of the Hilbert space that arises in thermofield dynamics method.
We shall determine the condensate function $f(\vec k)$,
and the functions $\theta_-(\vec k,\beta)$ and $\theta_+(\vec k,\beta)$ 
of the thermal vacuum through minimisation of the thermodynamic potential.
To evaluate the expectation value of the energy density
with respect to the thermal vacuum, we shall use the following formula.
\begin{equation}
\langle \psi_\gamma^\dagger (\vec x)\psi_\delta(\vec y)\rangle_\beta=
\frac{1}{(2\pi)^3}\int
\Big ( \Lambda _-(\vec k,\beta)\Big )_{\delta\gamma}
e^{-i\vec k .(\vec x-\vec y)}d\vec k,
\end{equation}
where
\begin{eqnarray}
\Lambda_-(\vec  k,\beta)&=&\frac{1}{2}\bigg[
(\cos^2\theta_++\sin^2\theta_-)
-\bigg(\gamma ^0\cos(\chi(\vec k) - 2 f(\vec k))\nonumber\\
&+&\vec\alpha \cdot \hat k~ \sin(\chi(\vec k)-2 f(\vec k))\bigg)
(\cos^2\theta_+-\sin^2\theta_-)\bigg].
\label{jjp}
\end{eqnarray}
We now proceed to calculate the energy density,
\begin{equation}
\epsilon\equiv\langle {\cal H}\rangle_\beta =\epsilon_N
+\epsilon_\sigma+\epsilon_\omega
\end{equation}
with 
\begin{mathletters}
\begin{equation}
\epsilon_N=-\frac{\gamma}{(2\pi)^3}\int d \vec k \Bigg[
\epsilon (\vec k)\cos 2f(\vec k)-\frac{g_\sigma \sigma_0}{\epsilon (k)}
\bigg(M \cos 2f(\vec k)+|\vec k|
\sin2 f (\vec k)\bigg)\Bigg] (\cos^2 \theta_+ -\sin^2 \theta _-)
\end{equation}
\begin{equation}
\epsilon_\sigma= \frac{1}{2}m_\sigma^2 \sigma_0^2,
\end{equation}
and 
\begin{equation}
\epsilon_\omega=g_\omega \omega_0 ~ \gamma (2\pi)^{- 3}\int d \vec k 
(\cos^2 \theta_+ +\sin^2 \theta _-)-\frac{1}{2}m_\omega^2 \omega_0^2.
\end{equation}
\end{mathletters}
The thermodynamic potential is then given as
\begin{equation}
\Omega=\epsilon-\frac{1}{\beta}{\cal S}-\mu \rho_B,
\label{thermpot}
\end{equation}
with the entropy density
\begin{eqnarray}
{\cal S}&=&-\gamma (2\pi)^{-3} \int d \vec k \Bigg[
\sin^2 \! \theta_- \ln(\sin^2\!\theta_-)+
\cos^2 \! \theta_- \ln(\cos^2\!\theta_-)\nonumber\\
&+&
\sin^2 \! \theta_+ \ln(\sin^2\!\theta_+)+
\cos^2 \! \theta_+ \ln(\cos^2\!\theta_+)
\Bigg]+{\cal S}_\sigma +{\cal S}_\omega
\label{entrf}
\end{eqnarray}
and the baryon density
\begin{equation}
\rho_B=\gamma (2\pi)^{- 3}\int d \vec k 
(\cos^2 \theta_+ +\sin^2 \theta _-).
\end{equation}
In the above, $\gamma$  is the spin  isospin degeneracy factor 
and is equal to $4$ for nuclear matter. Further, ${\cal S}_\sigma$ and
${\cal S}_\omega$ are the contributions to the entropy density from
$\sigma$- and $\omega$-mesons, respectively. It may be noted here that
these are independent of the functions $f(\vec k)$, $\theta_\pm(\vec k,\beta)$
associated with the nucleons and hence are not relevant for the nuclear
matter properties at zero temperature.
Extremising the thermodynamic potential $\Omega$ with respect
to the condensate function $f(\vec k)$ and the functions $\theta_{\mp}$ yields
\begin{equation}
\tan 2 f(\vec k)=\frac{g_\sigma \sigma_0 |\vec k|}{\epsilon(k)^2+
M g_\sigma \sigma_0}
\end{equation}
and
\begin{equation}
\sin^2 \theta _{\mp}=\frac{1}{\exp(\beta(\epsilon^*(k)\mp \mu^{*})) +1}
\end{equation}
with $\epsilon^*(k)=(k^2+{M^*}^2)^{1/2}$ and $\mu^{*}=\mu -g_\omega
\omega_0$ as the effective energy density and effective chemical
potential, where the effective nucleon mass is $M^{*}=M+g_\sigma \sigma_0$.

Then the expression for the energy density becomes
\begin{equation}
\epsilon=\epsilon_N+\epsilon_\sigma+\epsilon_\omega,
\end{equation}
with
\begin{mathletters}
\begin{equation}
\epsilon_N=\gamma (2\pi)^{-3}\int d \vec k (k^2+{M^*}^2)^{1/2} 
(\sin^2 \theta_- -\cos ^2 \theta_+) ,
\end{equation}
\begin{equation}
\epsilon_\sigma=\frac{1}{2}m_\sigma^2 \sigma_0^2,
\label{sgnc}
\end{equation}
and
\begin{equation}
\epsilon_\omega=
g_\omega \omega_0 ~\gamma (2\pi)^{-3}\int d \vec k 
(\sin^2 \theta_-+\cos^2 \theta_+) -\frac{1}{2}m_\omega^2 \omega_0^2.
\end{equation}
\end{mathletters}

We now proceed to study the properties of nuclear
matter at zero temperature. In that limit the distribution
functions for the baryons and antibaryons are given as
\begin{equation}
\sin^2\theta_-=\Theta(\mu^*-\epsilon^*(\vec k));\hspace {2cm} 
\sin^2\theta_+=0.
\end{equation}

The energy density after subtracting out the pure vacuum
contribution then becomes
\begin{eqnarray}
\epsilon_0& \equiv& \epsilon(\theta_-,f)-\epsilon(\theta_-=0,f=0)
\nonumber\\
&=& \epsilon_{MFT}+\Delta \epsilon
\label{enh}
\end{eqnarray}
with
\begin{equation}
\epsilon_{MFT}=\gamma (2\pi)^{-3}\int\limits_{|\vec k|<k_F} 
d \vec k(k^2+{M^*}^2)^{1/2}
+\frac{1}{2}m_\sigma^2 \sigma_0^2
+ g_\omega \omega_0 \rho_B -\frac{1}{2}m_\omega^2 \omega_0^2
\end{equation}
and
\begin{equation}
\Delta \epsilon= -\gamma (2\pi)^{-3}\int d \vec k \bigg[(k^2+{M^*}^2)^{1/2}-
(k^2+M^2)^{1/2}-\frac{g_\sigma \sigma_0 M}{(k^2+M^2)^{1/2}}\bigg].
\label{div}
\end{equation}
The above expression for the energy density is divergent.
It is renormalised \cite{chin} by adding the counter terms
\begin{equation}
\epsilon_{ct}=\sum_{n=1}^{4} C_n \sigma_0 ^n.
\label{ct}
\end{equation}
The addition of the counter term linear in $\sigma_0$ amounts to normal
ordering of the scalar density in the perturbative vacuum
and cancels exactly with the last term in equation (\ref{div})
\cite{chin}. The first two terms of the same equation corresponds
to the shift in the Dirac sea arising from the change in the nucleon
mass at finite density when $\sigma$ acquires a vacuum expectation
value, and consequent divergences cancel with the counter terms of 
(\ref{ct}) with higher powers in $\sigma_0$ \cite{chin}. 
Then we have the expression for the finite renormalised energy density
\begin{equation}
\epsilon_{ren}=
\epsilon_{MFT}+\Delta \epsilon_{ren},
\end{equation}
where
\begin{eqnarray}
\Delta \epsilon_{ren}&=& -\frac{\gamma}{16 \pi^2}\bigg[
{M^*}^4 \ln \bigl(\frac{M^{*}}{M}\bigr)+M^3 (M-{M^*})
-\frac{7}{2}M^2  (M-{M^*})^{2}\nonumber\\
 &+& \frac{13}{3} M  (M-{M^*})^3 
-\frac{25}{12} M  (M-{M^*})^4\bigg].
\label{rhf}
\end{eqnarray}
For a given baryon density  as given by
\begin{equation}
\rho_B=\gamma(2\pi)^{-3}\int d \vec k \Theta(k_F-k),
\end{equation}
the thermodynamic potential given by equation (\ref{thermpot})
is now finite and is a function of $\sigma_0$ and $\omega_0$.
This when minimised with respect to
$\sigma_0$ gives the self-consistency condition for the effective nucleon mass,
\begin{equation}
M^*=M-\frac{g_\sigma^2}{m_\sigma^2}\frac{\gamma}{(2\pi)^3}
\int d\vec k \frac{M^*}{\epsilon(k)^* }\Theta(k_F-k)+\Delta M^*
\end{equation}
where
\begin{equation}
\Delta M^*=\frac{g_\sigma^2}{m_\sigma^2}\frac{\gamma}{(2\pi)^3}
\bigg[ {M^*}^3 \ln \bigl(\frac{M^{*}}{M}\bigr)+M^2 (M-{M^*})
-\frac{5}{2}M^2  (M-{M^*})^{2} + \frac{11}{6} M  (M-{M^*})^3 \bigg]
\end{equation}

We note that the self-consistency condition for the
effective nucleon mass as well as the energy density as obtained here
through an explicit construct of a state with baryon-antibaryon
condensates are identical to those obtained through
summing tadpole diagrams for the baryon propagator
in the relativistic Hartree approximation \cite{chin}.

\section{Ansatz state with baryon antibaryon and sigma condensates}
We next consider the quantum corrections due to the
scalar mesons as arising from a vacuum realignment
with sigma condensates. This means that the $\sigma$-field is no
longer classical, but is now treated as a quantum field.
As will be seen later, a quartic term in the sigma field
would favour such condensates. Self-interactions
of scalar fields with cubic and quartic terms
have been considered earlier \cite{boguta}
in the no-sea approximation \cite{finnl} as well as including the
quantum  corrections arising from the sigma fields \cite{qhd,glen,fox}.
They may be regarded as mediating three- and four-body 
interactions between the nucleons.
The best fits to incompressibility in nuclear matter,
single-particle spectra and properties of deformed nuclei are
achieved with a negative value for the quartic coupling in the
sigma field. However, with such a negative coupling the energy spectrum
of the theory becomes unbounded from below \cite{furnstahl}
for large $\sigma$ and hence it is impossible to study
excited spectra or to include vacuum polarisation effects.

Including a quartic scalar self-interaction, eq. (\ref{lwsg}) is modified to 
\begin{equation}
{\cal H}_\sigma= \frac{1}{2}\partial_\mu \sigma \partial^\mu \sigma
+\frac{1}{2} m_\sigma ^2 \sigma^2+\lambda \sigma^4,
\label{lwsgm}
\end{equation}
with $m_\sigma$ and $\lambda$ being the bare mass and coupling 
constant respectively. The $\sigma$ field satisfies the quantum algebra
\begin{equation}
[\sigma (\vec x), {\dot \sigma} (\vec y)]=i \delta (\vec x-\vec y).
\label{gep2}
\end{equation}
We may expand the field operators in terms of creation and 
annihilation operators at time $t=0$ as
\begin{mathletters}
\begin{equation}
\sigma (\vec x,0)={1\over{(2 \pi)^{3/2}} }\int{{d\vec k\over{\sqrt
{2 \omega (\vec k)}}}\left(a(\vec k)+
a^\dagger(-\vec k)\right)e^{i\vec k\cdot\vec x}},
\end{equation}
\begin{equation}
\dot \sigma (\vec x,0)={i\over{(2 \pi)^{3/2}} }
\int {d\vec k\sqrt {\omega(\vec k)\over 2} \left(-a(\vec k)+
a^\dagger(-\vec k)\right)e^{i\vec k\cdot\vec x}}.
\end{equation}
\label{expan}
\end{mathletters}
In the above, $\omega (\vec k)$ is an arbitrary function which 
for free fields is given by $\omega (\vec k)=\sqrt{\vec k^2+m_\sigma^2}$ 
and the perturbative vacuum is defined corresponding to this basis through 
$a\mid vac\rangle=0$. The expansions (\ref{expan}) and the quantum
algebra (\ref{gep2}) yield the commutation relation for the operators
$a$ as
\begin{equation}
$$\bigl[a(\vec k), {a^\dagger(\vec k')}\bigr ]\,=\, \delta (\vec k-\vec k').
\label{gepq}
\end{equation}

As seen in the previous section a realignment of the
ground state from $\mid vac\rangle$ to $\mid vac'\rangle$
with nucleon condensates amounts to including quantum effects. We shall adopt a
similar procedure now to calculate the quantum corrections 
arising from the $\sigma$-field. We thus modify the ansatz for the trial ground 
state as given by (\ref{barcon}) to include $\sigma$ condensates as \cite{ph4}
\begin{equation}
|\Omega\rangle=U_\sigma U_F|vac\rangle,
\label{barsgcon}
\end{equation}
with
\begin{equation}
U_\sigma=U_{II}U_{I}
\label{gep5}
\end{equation}
where $U_{i}=\exp(B_i^\dagger~-~B_i),\,(i=I,II)$. Explicitly the
$B_{i}$ are given as
\begin{mathletters}
\begin{equation}
B_I^\dagger=\int {d\vec k \sqrt{\omega (\vec k)\over 2}
f_\sigma(\vec k) a^\dagger(\vec k)},
\end{equation}
and
\begin{equation}
{B_{II}}^\dagger={1\over 2}\int d\vec k g(\vec k){a'}^\dagger(\vec k)
{a'}^\dagger(-\vec k). 
\end{equation}
\end{mathletters}
In the above, $a'(\vec k)=U_I a(\vec k) U_I^{-1}=a(\vec k)-
\sqrt{\frac{\omega (\vec k)}{2}}f_\sigma(\vec k)$ corresponds to a
shifted field operator associated with the coherent state \cite{ph4} and 
satisfies the same quantum algebra as given in eq. (\ref {gepq}).
Thus in this construct for the ground state we have two
functions $f_\sigma(\vec k)$ and $g(\vec k)$ which will be determined through
minimisation of energy density. Further, since $\mid \Omega\rangle$ contains 
an arbitrary number of $a'^{\dagger}$ quanta, $a'\mid \Omega\rangle\,\not=\,0$. 
However, we can define the basis $b(\vec k)$, $b^\dagger(\vec  k)$ 
corresponding to $\mid \Omega\rangle$ through the Bogoliubov transformation as
\begin{eqnarray}
\left(
\begin{array}{c}
b(\vec k)\\ b^\dagger(-\vec k)
\end{array}
\right)
& = & U _{II}
\left(
\begin{array}{c}
a'(\vec k)\\ a'^\dagger(-\vec k)
\end{array}
\right)
U_{II}^{-1}=
\left(
\begin{array}{cc}
\cosh\! g
& -\sinh\! g \\ -\sinh\!g & \cosh\!g 
\end{array}
\right)
\left(
\begin{array}{c}
a'(\vec k)\\ a'^\dagger(-\vec k)
\end{array}
\right).
\label{gep7}
\end{eqnarray}
It is easy to check that $b(\vec k)\mid \Omega\rangle=0 $. Further, to preserve 
translational invariance $f_\sigma(\vec k)$ has to be  proportional to $\delta
(\vec k)$ and  we take $f_\sigma(\vec k)=\sigma _{0} (2\pi)^{3/2}
\delta (\vec k)$. $\sigma_0$ will correspond to a
classical field of the conventional approach \cite{ph4}.
We next calculate the expectation value of the Hamiltonian density
for the $\sigma$-meson given by equation(\ref{lwsgm}).
Using the transformations (\ref{gep7}) it is easy to evaluate that
\begin{mathletters}
\begin{equation}
\langle \Omega\mid \sigma \mid \Omega\rangle=\sigma _{0},
\label{gep9a}
\end{equation}
but,
\begin{equation}
\langle \Omega\mid \sigma^2 \mid \Omega\rangle={\sigma _{0}}^{2}+I,
\label{gep9b}
\end{equation}
where
\begin{equation}
I={1 \over (2 \pi)^3}\int{{d\vec k \over {2\;\omega (k)}}
(\cosh \!2g +\sinh\!2g)}.
\label{gep9c}
\end{equation}
\end{mathletters}
Using equations (\ref{lwsgm}) and (44) the energy density of 
${\cal H}_\sigma$ with respect to the trial state becomes \cite{ph4}
\begin{eqnarray}
\epsilon_\sigma\equiv\langle \Omega\mid{\cal H}_\sigma\mid \Omega\rangle &=&
\frac{1}{2}{1 \over {(2 \pi)^3}} \int{d \vec k
\over 2\omega (k)}\Bigg[ k^{2}(\sinh\!2g +\cosh\!2g)
+\omega^2 (k)(\cosh\!2g -\sinh\!2g)\Bigg]
\nonumber\\
&+& \frac{1}{2}m_\sigma^2I+6\lambda\sigma_0^2 I+3\lambda I^2
+\frac{1}{2}m_\sigma^2\sigma_0^2 +\lambda \sigma_0^4.
\label{en}
\end{eqnarray}
Extremising the above energy density with respect to the function $g(k)$ yields 
\begin{equation}
\tanh\!{2 g(k)}=-\,{{6 \lambda I+6 \lambda {\sigma _0}^2}\over {
{\omega (k)}^{2}+6 \lambda I+6 \lambda {\sigma _{0}}^{2}}}.
\label{gk}
\end{equation}
It is clear from the above equation that in the absence of a 
quartic coupling no such condensates are favoured since the
condensate function vanishes for $\lambda=0$.
Now substituting this value of $g(k)$ in the expression
for the $\sigma$-meson energy density yields 
\begin{equation}
\epsilon_\sigma=
{1 \over {2}}m_\sigma^2{\sigma _0}^2\,+\lambda {\sigma _0}^4
+\frac{1}{2}\frac{1}{(2\pi)^3}\int d\vec k (k^2+M_\sigma^2)^{1/2}
-3 \lambda I^2
\label{pot}
\end{equation}
where 
\begin{equation}
M_\sigma^2=m_\sigma^2+12\lambda I +12\lambda \sigma_0^2
\label{m2}
\end{equation}
with
\begin{equation}
I=\frac{1}{(2\pi)^3}\int\frac{d\vec k}{2} 
\frac{1}{(\vec k^2+M_\sigma^2)^{1/2}}
\label{I}
\end{equation}
obtained from equation (\ref{gep9c}) after substituting for the condensate 
function $g(k)$ as in equation (\ref{gk}).
The expression for the ``effective potential" $\epsilon_\sigma$
contains divergent integrals. Since our approximation is 
nonperturbatively self-consistent, the field-dependent
effective mass $M_\sigma$ is also not well defined because
of the infinities in the integral $I$ given by equation (\ref{I}).
Therefore we first obtain a well-defined finite expression
for $M_\sigma$ by renormalisation. We use the
renormalisation prescription of ref. \cite{politzer}
and thus obtain the renormalised mass $m_R$ and coupling $\lambda_R$ through
\begin{mathletters}
\begin{equation}
\frac{m_R^2}{\lambda_R}=
\frac{m^2}{\lambda}+12I_1(\Lambda),
\label{mr}
\end{equation}
\begin{equation}
\frac{1}{\lambda_R}=
\frac{1}{\lambda}+12I_2(\Lambda , \mu),
\label{lr}
\end{equation}
\end{mathletters}
where $I_1$ and $I_2$ are the integrals
\begin{mathletters}
\begin{equation}
I_1(\Lambda)=\frac{1}{(2\pi)^3}\int\limits_{|\vec k|<\Lambda} 
\frac{d \vec k}{2k},
\end{equation}
\begin{equation}
I_2(\Lambda ,\mu)=\frac{1}{\mu^2}\int\limits_{|\vec k|<\Lambda} 
\frac{d \vec k}{(2\pi)^3}\Big (\frac{1}{2k}
-\frac{1}{2\sqrt{k^2+\mu^2}}\Big),
\label{i2}
\end{equation}
\end{mathletters}
with $\mu$ as the renormalisation scale and
$\Lambda$ as an ultraviolet momentum cut-off. It may be noted here that
with the use of the above renormalisation prescription the effective
sigma mass $M_\sigma$ and the energy density ultimately become
independent of $\Lambda$ and stay finite in the limit $\Lambda \rightarrow
\infty$. Using equations (\ref{mr}) and 
(\ref{lr}) in equation (\ref{m2}), we have
the gap equation for $M_\sigma^2$ in terms of the renormalised parameters as
\begin{equation}
M_\sigma^2=m_R^2+12\lambda_R\sigma_0^2+12\lambda_R I_f(M_\sigma),
\label{mm2}
\end{equation}
where
\begin{equation}
I_f(M_\sigma)=\frac{M_\sigma^2}{16\pi^2}\ln \Big(\frac{M_\sigma^2}{\mu^2} \Big).
\label{if}
\end{equation}
Then using the above equations  we simplify equation (\ref{pot}) to obtain the
energy density for the $\sigma$ in terms of $\sigma_0$ as
\begin{equation}
\epsilon_\sigma=3\lambda_R\Big(\sigma_0^2+\frac{m_R^2}{12\lambda_R}\Big)^2
+\frac {M_\sigma^4}{64\pi^2}\Biggl(\ln\Big(\frac{M_\sigma^2}{\mu^2}\Big)
-\frac{1}{2} \Biggr)
-3\lambda_R I_f^2-2\lambda\sigma_0^4.
\label{vph}
\end{equation}
The above expression is given in terms of the renormalized 
$\sigma$ mass $m_R$ and the renormalized coupling $\lambda_R$ except for the
last term which is still in terms of the bare coupling constant $\lambda$
and did not get renormalised because of the structure of the gap
equation \cite{pi}.
However, from the renormalisation condition (\ref{lr}) 
it is easy to see that when $\lambda_R$ is kept fixed, 
as the ultraviolet cut-off $\Lambda$ in eq. (\ref{i2})
goes to infinity, the bare coupling $\lambda \rightarrow 0_-$.
Therefore the last term in eq. (\ref{vph}) will be neglected
in the numerical calculations.

After subtracting the vacuum contribution, we get
\begin{eqnarray}
\Delta \epsilon_\sigma &=& \epsilon_\sigma-\epsilon_\sigma(\sigma_0=0)
\nonumber \\
&=& \frac{1}{2} m_R^2 \sigma_0^2+ 3\lambda_R \sigma_0^4 
+\frac {M_\sigma^4}{64\pi^2}
\Biggl(\ln\Big(\frac{M_\sigma^2}{\mu^2}\Big)-\frac{1}{2} \Biggr)
-3\lambda_R I_f^2\nonumber\\
&-&\frac {M^4_{\sigma,0}}{64\pi^2}
\Biggl(\ln\Big(\frac{M_{\sigma,0}^2}{\mu^2}\Big)-\frac{1}{2} \Biggr)
+3\lambda_R I_{f0}^2,
\label{vph0}
\end{eqnarray}
where $M_{\sigma,0}$ and $I_{f0}$ are the expressions
as given by eqs. (\ref{mm2}) and (\ref{if}) with $\sigma_0=0$.

In the limit of the coupling, $\lambda_R=0$, one can see that
eq. (\ref{vph0}) reduces to eq. (\ref{sgnc}) as it should.
Also, we note that the sign of $\lambda_R$ must be chosen
to be positive, because otherwise the energy density would become 
unbounded from below with vacuum fluctuations \cite{serot2,fox,furnstahl}.

The expectation value for the energy density after subtracting out
the vacuum contribution as given by eq. (\ref{enh}),
now with sigma condensates is modified to
\begin{equation}
\epsilon_0=\epsilon_0^ {finite}+\Delta\epsilon,
\end{equation}
where
\begin{equation}
\epsilon_0 ^{finite}=\gamma (2\pi)^{-3}\int\limits_{|\vec k|<k_F} 
d\vec k (k^2+{M^*}^2)^{1/2} + g_\omega \omega_0 \rho_B
-\frac{1}{2}m_\omega^2 \omega_0^2 +\Delta\epsilon_\sigma,
\end{equation}
with $\Delta\epsilon_\sigma$ given through eq. (\ref{vph0})
and $\Delta\epsilon$ is the divergent part of the energy density given 
by equation (\ref{div}). We renormalise by adding the same counter 
terms as given by (\ref{ct}) so that as earlier the renormalised mass and
the renormalised quartic coupling remain unchanged \cite{qhd}. 
This yields the expression for the energy density
\begin{equation}
\epsilon_{ren}=\epsilon_0^ {finite}+\Delta\epsilon_{ren},
\label{enf}
\end{equation}
with $\Delta\epsilon_{ren}$ given by eq. (\ref{rhf}).
As earlier the energy density is to be minimised with respect to
$\sigma_0$ to obtain the optimised value for $\sigma_0$,
thus determining the effective mass $M^*$  in a self-consistent manner.

The energy density from the $\sigma$ field as given by eq.
(\ref{vph0}) is still in terms of the renormalisation scale $\mu$
which is arbitrary. We choose this to be equal to 
the renormalised sigma mass $m_R$ in doing the numerical
calculations. This is because changing $\mu$ would mean changing 
the quartic coupling $\lambda_R$, and $\lambda_R$ here
enters as a parameter to be chosen to give the incompressibility
in the correct range.

For a given baryon density, $\rho_B$, the binding energy for nuclear matter is

\begin{equation}
E_B=\epsilon_{ren}/\rho_B-M,
\label{be}
\end{equation}

The  parameters $g_\sigma$, $g_\omega$ and $\lambda_R$ are fitted
so as to describe the ground-state properties of nuclear matter
correctly. We discuss the results in the next section.

\section{Results and Discussions}

We now proceed with the numerical calculations to study the nuclear 
matter properties at zero temperature.
We take the nucleon and $\omega$-meson masses to be their experimental
values as 939 MeV and 783 MeV. 
We first calculate the binding energy per nucleon as given in equation
(\ref{be}) and fit the scalar and vector couplings $g_\sigma$ and $g_\omega$
to get the correct saturation properties of nuclear matter. This
involves first minimising the energy density in eq. (\ref{enf}) 
with respect to $\sigma_0$ to get the optimised scalar field 
ground state expectation
value $\sigma_{min}$. This procedure also naturally includes obtaining the
in-medium $\sigma$-meson mass $M_\sigma$ through solving the gap equation
(\ref{mm2}) in a self-consistent manner. Obtaining the optimized 
$\sigma_{min}$
amounts to getting the effective nucleon mass $M^*=M+g_\sigma\sigma_{min}$. 
We fix the meson couplings from the saturation properties of the 
nuclear matter for given renormalised sigma mass and coupling $m_R$ 
and $\lambda_R$. Taking $m_R=520$ MeV,
the values of $g_\sigma$ and $g_\omega$ are $7.34$ and $8.21$ for
$\lambda_R=1.8$, and are $6.67$ and $7.08$ for $\lambda_R=5$ respectively. 
Using these values, we calculate the binding energy for nuclear 
matter as a function of the Fermi momentum and plot it in fig.~1.
In the same figure we also plot the results for the relativistic
Hartree and for the no-sea approximation. Clearly, including baryon and 
$\sigma$-meson quantum corrections
leads to a softer equation of state and the softening increases for
higher values of $\lambda_R$. The 
incompressibility of the nuclear matter is given as \cite{compres}
\begin{equation}
K=k_F^2 \frac{\partial^2 \epsilon}{\partial k_F^2}
\end{equation}
evaluated at the saturation Fermi momentum. The value of $K$
is found to be $401$ MeV for $\lambda_R=1.8$ and $329$ MeV for $\lambda_R=5$.
These are smaller than the mean-field result
of 545 MeV \cite{walecka}, as well as that of relativistic Hartree of 450 MeV
\cite{chin}  and are similar to those obtained 
in ref. \cite{fox} containing cubic and quartic self-interaction of the 
$\sigma$-meson.

In fig.~2, we plot the effective nucleon mass $M^*=M+g_\sigma
\sigma_{min}$ as a function of Fermi momentum 
with $\sigma_{min}$ obtained from the minimisation of 
the energy density in a self-consistent
manner. At the saturation density of $k_F=1.42$ fm$^{-1}$, we get
$M^*=0.752M$ and $0.815M$ for $\lambda_R=1.8$ and $\lambda_R=5$, respectively.
These values may be compared with the results of $M^*= 0.56M$ in the 
no-sea approximation and of $0.72M$ in the relativistic Hartree.

In fig.~3, we plot the vector and the scalar potentials as functions
of $k_F$ for sigma self coupling $\lambda_R=1.8$ and 5. 
At saturation density the scalar and vector contributions are
$U_S\equiv g_\sigma\sigma_{min}=-232.7$ MeV 
and $U_V\equiv g_\omega\omega_0=163.4$ MeV for $\lambda_R=1.8$ and
are $-173.14$ MeV and $107.74$ MeV for $\lambda_R=5$ respectively. 
These give rise to the nucleon potential $(U_S+U_V)$ of $-69.3$ 
MeV and $-65.4$ MeV
and an antinucleon potential $(U_S-U_V)$ of $-396.1$ MeV and $-280.9$ MeV for 
$\lambda_R=1.8$ and 5. Clearly the inclusion of the quantum corrections
reduces the antinucleon potential as compared to both the relativistic 
Hartree ($-450$ MeV) \cite{chin} and the no-sea results ($-746$ MeV)
\cite{walecka}.

In fig.~4, we plot the in-medium $\sigma$-meson mass $M_\sigma$
of eq. (\ref{mm2}) as a function of baryon density for
$\lambda_R=1.8$ and $5$. $M_\sigma$ increases
with density as $\lambda_R$ is positive and the magnitude of $\sigma_{min}$ 
increases with density too. However, the change in $M_\sigma$ is rather small.

In fig.~5, we plot the incompressibility $K$ as a function of the quartic
coupling $\lambda_R$ for different values of $m_R$, the renormalised 
sigma mass in vacuum. The value of $K$ decreases with increase in
$\lambda_R$ similar to the results obtained in ref. \cite{fox}. 
In fig.~6, we plot the effective nucleon mass versus the sigma self
coupling for various values of $m_R$. The value of $M^*$ increases
with $\lambda_R$, which is a reflection of the diminishing nucleon-sigma
coupling strength for larger values of the quartic self-interaction.

To summarise we have used a nonperturbative approach to include quantum 
effects in nuclear matter using the framework of QHD.
Instead of going through a loop expansion and 
summing over an infinite series of Feynman diagrams we have included the
quantum corrections through a realignment of the ground state with baryon
as well as meson condensates. It is interesting to note that inclusion of 
baryon-antibaryon condensates with the particular ansatz determined through
minimisation of the thermodynamic potential yields the same results as
obtained in the relativistic Hartree approximation. This results in
a softer equation of state as compared to the no-sea approximation. 
Inclusion of scalar meson quantum corrections in a self-consistent 
manner leads to a further softening of the equation of state. 
The value for the incompressibility
of nuclear matter is within the range of 200--400 MeV \cite{kuono}.
It is known that most of the parameter sets which explain the
ground state properties of nuclear matter and finite nuclei quite well 
are with a negative quartic coupling. But
the energy spectrum in such a case is unbounded from below \cite{furnstahl}
for large $\sigma$ thus making it impossible to include
vacuum polarisation effects. We have included the quantum 
effects with a quartic self interaction
through sigma condensates taking the coupling to be positive. 
We have also calculated the effective mass of the sigma field
as modified by the quantum corrections from baryon and sigma fields.
The effective sigma mass is seen to increase with density. 

We have also looked at the behaviour of the incompressibility
as a function of the coupling $\lambda_R$ for various values of
sigma mass, which is seen to decrease with the
coupling. Finally, we have looked at the effect of the sigma 
quartic  coupling on the effective nucleon mass which grows
with the coupling. Generally, higher values of the quartic
term in the potential of the $\sigma$-meson tend to reduce the
large meson fields and thus the strong relativistic effects in the nucleon 
sector. Clearly, the approximation here lies in the specific ansatz 
for the ground-state structure. However, a systematic inclusion 
of more general condensates than the pairing one as used here might 
be an improvement over the present one. The method 
can also be generalised to finite temperature as well as 
to finite nuclei, e.g., using the local density approximation. 
Work in this direction is in progress.

\section{Acknowledgements}

The authors would like to thank J.M. Eisenberg
for extensive discussions, many useful suggestions and for a 
critical reading of the manuscript. AM and PKP acknowledge discussions with
H. Mishra during the work. They would also like to thank the 
Alexander von Humboldt-Stiftung for financial support and the 
Institut f\"ur Theoretische Physik for hospitality.

\vfil
\eject
\centerline{\bf Figures Captions}

\noindent Fig.~1. The binding energy for nuclear matter
as a function of Fermi momentum $k_F$ corresponding to
the no-sea and the relativistic Hartree approximations 
and the approach including quantum corrections
from baryon and $\sigma$-meson given by eq. (59).
It is seen that the equation of state is softer with such
quantum corrections.  \hfil\break

\noindent Fig.~2. The effective nucleon mass for nuclear matter as a 
function of the Fermi momentum, $k_F$.  \hfil\break

\noindent Fig.~3. The scalar potential $U_S$ (negative values) 
and vector potential $U_V$  (positive values)
for nuclear matter as functions of Fermi momentum $k_F$.\hfil\break

\noindent Fig.~4. The in-medium $\sigma$-meson mass $M_\sigma$
of eq. (\ref{mm2}) as a function of density, which is seen to
increase with density. However, the change is seen
to be rather small.\hfil\break

\noindent Fig.~5. The incompressibility $K$ versus the quartic coupling
$\lambda_R$ for various values of $m_R$, which is
seen to decrease with $\lambda_R$. The values of $K$ are higher for
larger values of the sigma mass.\hfil\break

\noindent Fig.~6. The effective nucleon mass as a function of $\lambda_R$
for different values of $m_R$. It is seen to decrease with increase
in the coupling.
\vfil
\eject

\begin{figure}
\epsfbox{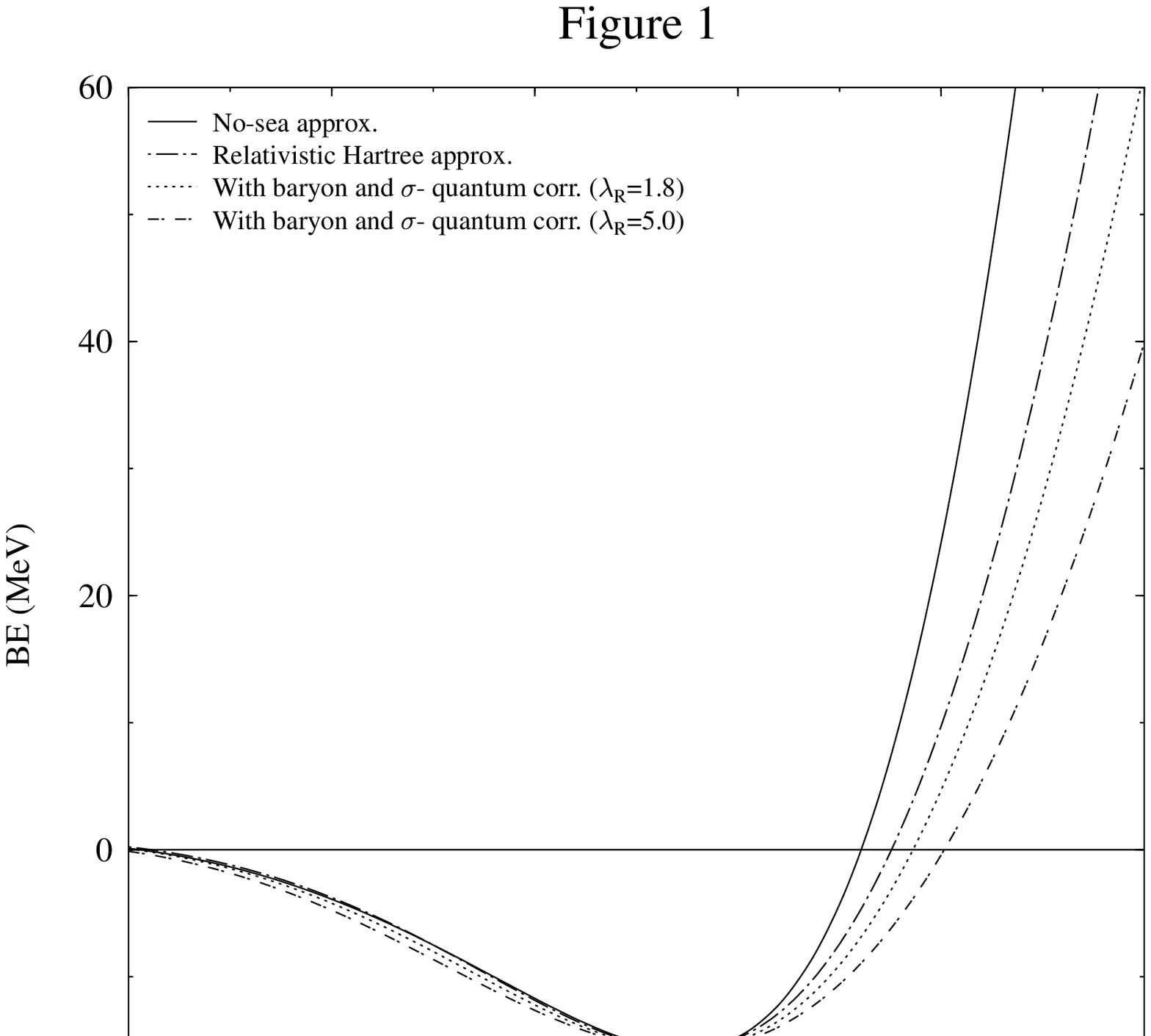}
\end{figure}
\begin{figure}
\epsfbox{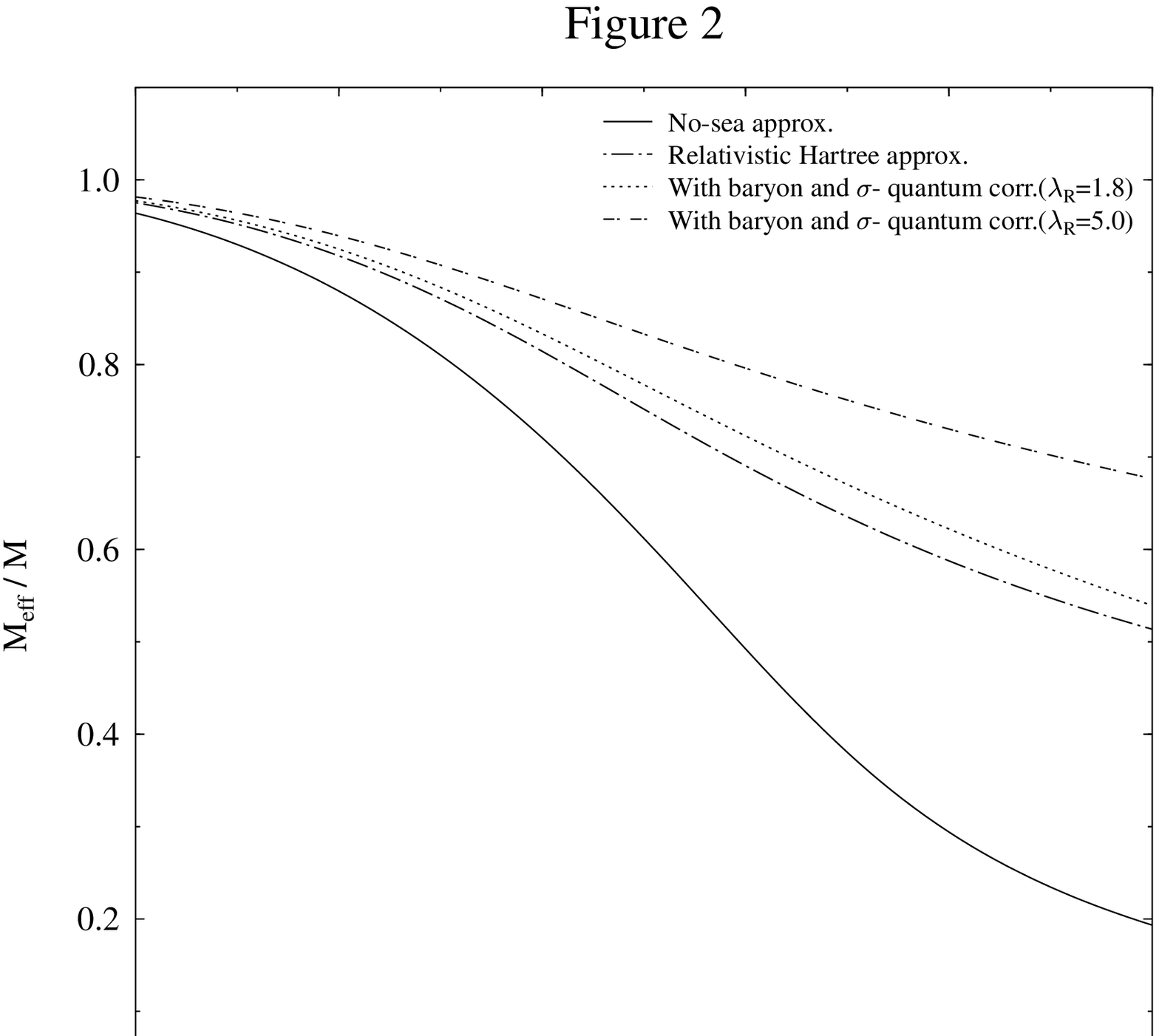}
\end{figure}
\begin{figure}
\epsfbox{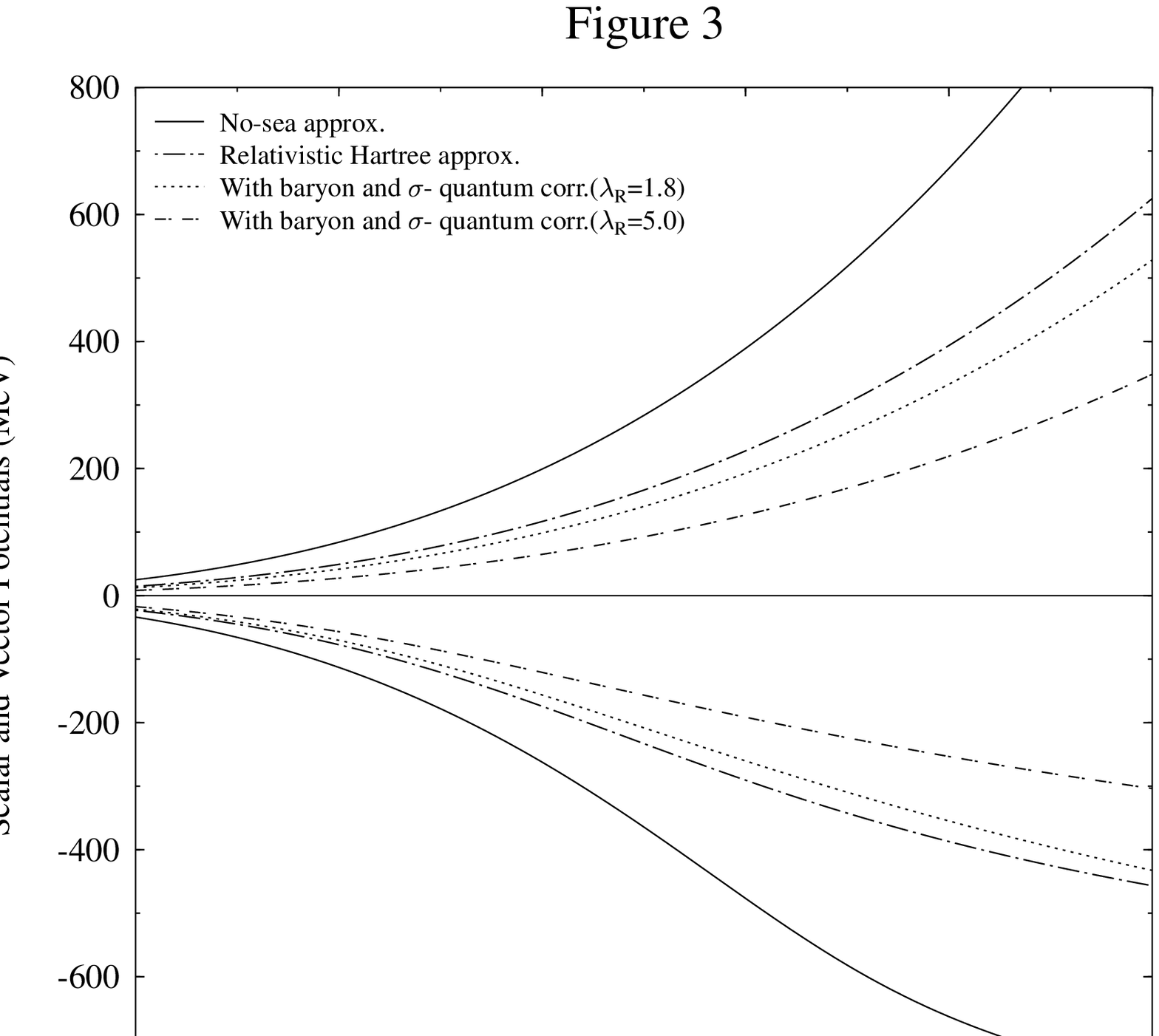}
\end{figure}
\begin{figure}
\epsfbox{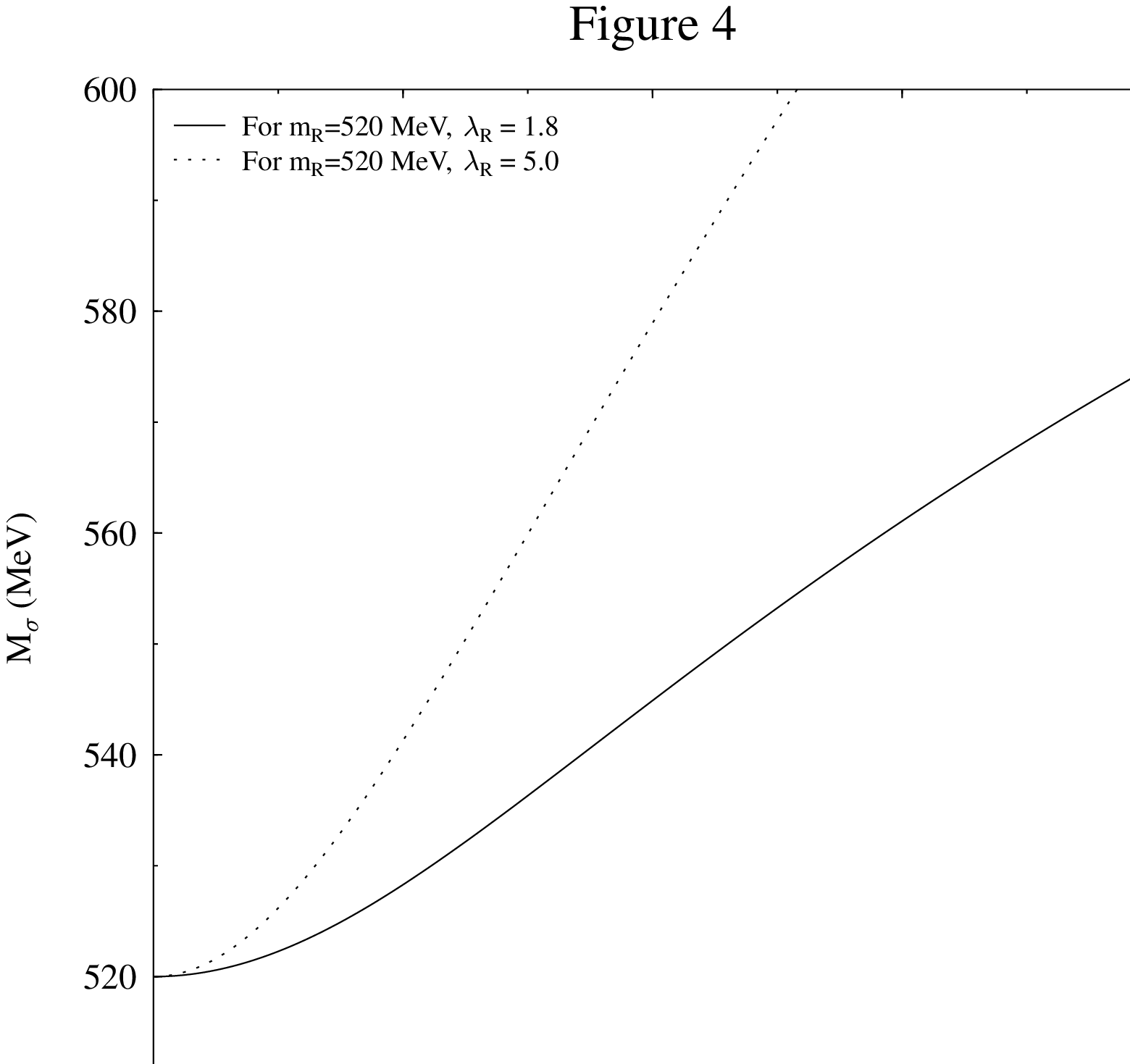}
\end{figure}
\begin{figure}
\epsfbox{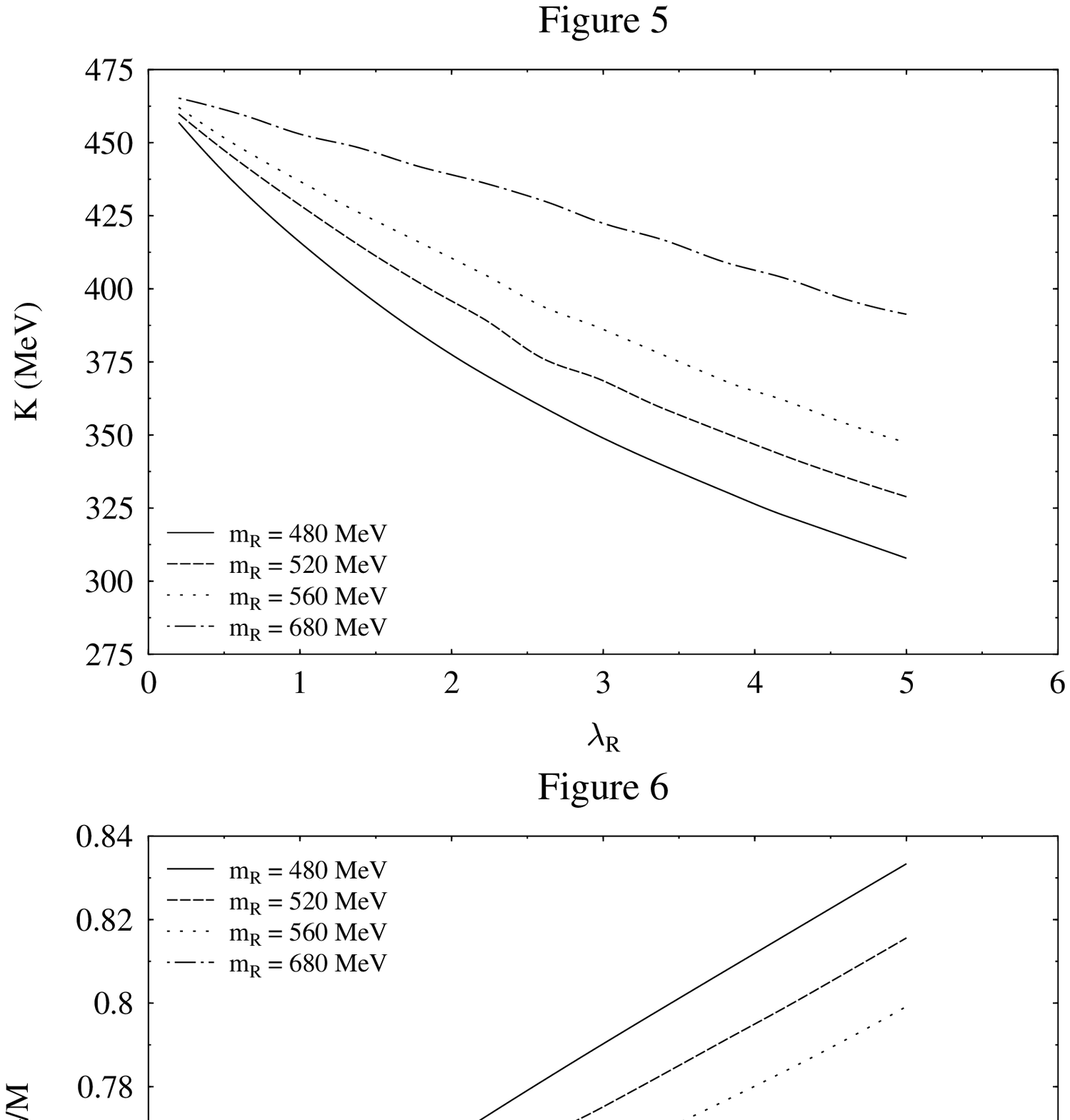}
\end{figure}

\begin{thebibliography}{99}
\bibitem{qhd}B.D. Serot and J.D. Walecka, Adv. Nucl. Phys. {\bf 16}, 1 (1986);
B.D. Serot, J.D. Walecka, nucl-th/9701058, to appear in Int. J. Mod. Phys. E.
\bibitem{reinhard}P.-G. Reinhard, Rep. Prog. Phys. {\bf 52}, 439 (1989).
\bibitem{serot2}B.D. Serot, Rep. Prog. Phys. {\bf 55}, 1855 (1992).
\bibitem{walecka}J.D. Walecka, Ann. of Phys. {\bf 83}, 491 (1974).
\bibitem{nm}B.M. Waldhauser, J. Maruhn, H. St\"ocker and W. Greiner, Phys. Rev.
{\bf C38}, 1003 (1988).
\bibitem{finnl}M. Rufa, P.G. Reinhard, J. Maruhn, W. Greiner,
Phys. Rev. {\bf C38}, 390 (1988); Y.K. Gambhir, P. Ring, A. Thimet,
Ann. of Phys. {\bf 198}, 132 (1990); Y. Sugahara, H. Toki, 
Nucl. Phys. {\bf A574}, 557 (1994).
\bibitem{chin}S.A. Chin and J.D. Walecka, Phys. Lett. {\bf B 52}, 24 (1974);
S.A. Chin, Ann. of Phys. {\bf 108}, 301 (1977); R.J. Perry, Phys. Lett.
{\bf B 199}, 489 (1987).
\bibitem{twoloop} C.J. Horowitz and B.D. Serot, Phys. Lett. 
{\bf B 140}, 181 (1984);
R.J. Furnstahl, R.J. Perry and B.D. Serot, Phys. Rev. {\bf C40}, 321 (1989).
\bibitem{prakash} M. Prakash, P.J. Ellis and J.I. Kapusta, Phys. Rev.
{\bf C45}, 2518 (1992); R. Friedrich, K. Wehrberger and F. Beck,
Phys. Rev. {\bf C46}, 188 (1992).
\bibitem{horowitz} R.J. Furnstahl and C.J. Horowitz, Nucl. Phys.
{\bf A485}, 632, (1988).
Rev. {\bf C51}, 1754 (1995).
\bibitem{allen} M.P. Allendes and B.D. Serot, Phys. Rev. {\bf C45},
2975 (1992); B.D. Serot and H. Tang, Phys. Rev. {\bf C51}, 969
(1995).
\bibitem{cond} A. Mishra, H. Mishra, S.P. Misra and S.N. Nayak,
Pramana (J. of  Phys.) {\bf 37}, 59 (1991); {\em ibid}, Z. Phys. {\bf C57}, 
233 (1993); A. Mishra, H. Mishra, V. Sheel, S.P. Misra and P. K. Panda,
Int. J. Mod. Phys. {\bf E3}, 93 (1996).
\bibitem{ph4} A. Mishra and H. Mishra, hep-ph/9611365, to appear in J. Phys. G.
\bibitem{gn} H. Mishra, S.P. Misra and A. Mishra, Int. J. Mod. Phys. 
{\bf A3}, 2331 (1988);M.G. Mitchard, A.C. Davis and A.J. Macfarlane,
Nucl.Phys {\bf B325}, 470 (1989).
\bibitem{chirl}A. Mishra, H. Mishra and S.P. Misra, Z. Phys. {\bf C59}, 
159 (1993).
\bibitem{prd}S.P. Misra, Phys. Rev. {\bf D18}, 1661 (1978).
\bibitem{tfd} H. Umezawa, H. Matsumoto and M. Tachiki,
{\em Thermofield Dynamics and Condensed States} 
(North-Holland, Amsterdam, 1982);
P.A. Henning, Phys. Rep. {\bf 253}, 235 (1995).
\bibitem{boguta}J. Boguta and A.R. Bodmer, Nucl. Phys. {\bf A 292}, 413 (1977).
\bibitem{glen}N.K. Glendenning, Nucl. Phys. {\bf A493}, 521 (1989).
\bibitem{fox} W.R. Fox, Nucl. Phys. {\bf A 495}, 463 (1989);
ibid, Ph. D. thesis, Indiana University.
\bibitem{furnstahl}R.J. Furnstahl, C.E. Price and G.E. Walker,
Phys. Rev. {\bf C36}, 2590 (1987).
\bibitem{politzer} S. Coleman, R. Jackiw and H.D. Politzer,
Phys. Rev. {\bf D10}, 2491 (1974).
\bibitem{pi} S.Y. Pi and M. Samiullah, Phys. Rev. {\bf D36}, 3121 (1987);
G. A. Camelia and S.Y. Pi, Phys. Rev. {\bf D47}, 2356 (1993).
\bibitem{compres} J.P. Blaizot, D. Gogny and B. Grammaticos,
Nucl. Phys. {\bf A265}, 315 (1975).
\bibitem{kuono}H. Kuono, N. Kakuta, N. Noda, T. Mitsumori, A. Hasegawa, Phys.
Rev. {\bf C51}, 1754 (1995).
\end{thebibliography}
\end{document}